\documentclass[aps,pra,twocolumn,superscriptaddress]{revtex4-2}

\usepackage{color}
\usepackage{xcolor}
\usepackage{array}
\usepackage{amsmath}
\usepackage{amssymb}
\usepackage{amsfonts}
\usepackage{algorithm}
\usepackage{algorithmicx}
\usepackage{algpseudocode}
\usepackage{graphicx}
\usepackage{subfigure}
\usepackage{textcomp}
\usepackage{hyperref}
\hypersetup{
	colorlinks=true,
	pdfstartview=Fit,
	breaklinks=true}
\usepackage[braket, qm]{qcircuit}

\usepackage{algpseudocode}

\begin{document}


\title{Quantum-Based Self-Attention Mechanism for Hardware-Aware Differentiable Quantum Architecture Search}


\author{Yuxiang Liu}
\thanks{These authors contributed equally to this work.}
\affiliation{National Mobile Communications Research Laboratory, Southeast University, Nanjing 210096, China}
\affiliation{Purple Mountain Laboratories, Nanjing 211111, China}
\affiliation{Frontiers Science Center for Mobile Information Communication and Security, Southeast University, Nanjing 210096, China}

\author{Sixuan Li}
\thanks{These authors contributed equally to this work.}
\affiliation{Frontiers Science Center for Mobile Information Communication and Security, Southeast University, Nanjing 210096, China}
\affiliation{State Key Lab of Millimeter Waves, Southeast University, Nanjing 211189, China}

\author{Fanxu Meng}
\affiliation{College of Artificial Intelligence, Nanjing Tech University, Nanjing 211800, China}

\author{Zaichen Zhang}
\email{zczhang@seu.edu.cn}
\affiliation{National Mobile Communications Research Laboratory, Southeast University, Nanjing 210096, China}
\affiliation{Purple Mountain Laboratories, Nanjing 211111, China}
\affiliation{Frontiers Science Center for Mobile Information Communication and Security, Southeast University, Nanjing 210096, China}

\author{Xutao Yu}
\email{yuxutao@seu.edu.cn}
\affiliation{Purple Mountain Laboratories, Nanjing 211111, China}
\affiliation{Frontiers Science Center for Mobile Information Communication and Security, Southeast University, Nanjing 210096, China}
\affiliation{State Key Lab of Millimeter Waves, Southeast University, Nanjing 211189, China}


\date{\today}

\begin{abstract}
The automated design of parameterized quantum circuits for variational algorithms in the Noisy Intermediate-Scale Quantum (NISQ) era faces a fundamental limitation, as conventional differentiable architecture search relies on classical models that fail to adequately represent quantum gate interactions under hardware noise. We introduce the Quantum-Based Self-Attention for Differentiable Quantum Architecture Search (QBSA-DQAS), a meta-learning framework featuring Quantum-Based self-attention and hardware-aware multi-objective search for automated architecture discovery.
The framework employs a two-stage Quantum-Based self-attention module. First, it computes contextual dependencies by mapping architectural parameters through parameterized quantum circuits, extracting high-dimensional feature representations that replace classical similarity metrics with quantum-derived attention scores. Second, it applies position-wise quantum transformations for feature enrichment. Architecture search is guided by a task-agnostic multi-objective function jointly optimizing noisy expressibility and Probability of Successful Trials (PST). A post search optimization stage applies gate commutation, fusion, and elimination to reduce circuit complexity.
Experimental validation demonstrates superior performance on Variational Quantum Eigensolver (VQE) tasks and large-scale Wireless Sensor Network (WSN). For VQE on H$_2$, QBSA-DQAS achieves 0.9 accuracy compared to 0.89 for standard DQAS, outperforming classical attention baselines. Post-search optimization via deterministic simplification rules reduces the 
complexity of the discovered circuits by up to 44\% in gate count and 47\% 
in depth without accuracy degradation. The framework maintains robust performance across three molecules and five IBM quantum hardware noise models. For WSN routing, discovered circuits achieve 8.6\% energy reduction versus QAOA and 40.7\% versus classical greedy method. These results establish the effectiveness of quantum-native architecture search for NISQ applications.
\end{abstract}


\maketitle

\section{Introduction}
Quantum computing is a rapidly advancing field with the potential to transform diverse domains, from fundamental science to machine learning. In recent years, significant progress has been demonstrated in areas such as image classification, drug discovery, and the solution of combinatorial optimization problems\cite{9643516, WOS:000753864600001,Cerezo2021NRP,QAOA_Zhou2020_Quantum,2021Quantum}. However, the current era of Noisy Intermediate-Scale Quantum (NISQ) computing imposes hardware constraints, including high noise levels, limited qubit counts, and short coherence times\cite{Burnett_2019}, which present a primary obstacle to realizing practical quantum advantage\cite{arute2019quantum,WOS:000819150300001,Bharti_2022,2024Current}. Within these constraints, Variational Quantum Algorithms (VQAs) have emerged as a leading computational paradigm\cite{WOS:000340611600001,Cerezo2021NRP}, employing a hybrid quantum-classical approach to solve complex optimization problems. However, the success of any VQA critically depends on the architecture of its parameterized quantum circuit (PQC) or ansatz. A well-designed ansatz is a key determinant of performance\cite{WOS:000613950500002}, as it must be sufficiently expressive to represent the solution to the target problem, yet compact enough for reliable execution on noisy quantum near-term hardware\cite{Cerezo_2021,McClean2018BarrenPlateaus,Kandala2017HEA}.

To address the challenge of ansatz design, the field has moved from manual methods towards automated Quantum Architecture Search (QAS). While early QAS strategies, based on evolutionary or reinforcement learning, demonstrated potential, they were often hampered by excessive computational costs in searching large circuit spaces. More recently, Differentiable Quantum Architecture Search (DQAS) has emerged as a more efficient paradigm\cite{WOS:000843298500001,QuantumNAS2021,liu2025outputpredictionquantumcircuits}. By relaxing the discrete choices of quantum gates into a continuous, differentiable search space, DQAS can leverage gradient-based optimization methods to explore vast architectural landscapes with significantly improved sample efficiency.

Despite its sample efficiency, the conventional DQAS framework suffers from a fundamental limitation. Its core mechanism relies on classical models to represent and evaluate the relationships between quantum operations, often using classical vectors and similarity metrics like the dot product. This creates an inherent representational mismatch, where a classical model is tasked with evaluating the relationships within a quantum system, whose complex, non-linear interactions are governed by the high-dimensional rules of quantum mechanics\cite{Schuld2019FeatureHilbert}. This incongruity is exacerbated by hardware noise, which non-trivially alters the behavior of quantum operations\cite{Wang2021NoiseInducedBP}. A similarity metric computed within an idealized, noise-free classical model offers little guidance for discovering an architecture that will be resilient on real, noisy hardware. Therefore, an effective QAS framework necessitates a search mechanism that is both intrinsically quantum in its operation and explicitly aware of hardware.

To resolve this incongruity, we reframe the challenge of QAS within the paradigm of meta-learning\cite{finn2017modelagnosticmetalearningfastadaptation}, or learning to learn. Our goal is not merely to solve a single problem, but to develop a framework that learns a universal strategy for designing effective circuits. We turn to the field of Quantum Machine Learning (QML) for a more suitable mechanism\cite{WOS:000457936400001}. We propose the Quantum-Based Self-Attention for Differentiable Quantum Architecture Search (QBSA-DQAS) framework. Our approach is inspired by quantum feature mapping techniques, a QML technique that leverages the exponentially large Hilbert space of a quantum system as a feature space\cite{WOS:000463851300001,WOS:000536095000002}. Our approach integrates a quantum-native search mechanism with a hardware-aware search objective. The search mechanism is a Quantum-Based Self-Attention (QBSA) module, which adopts the self-attention framework to capture global, contextual dependencies within a circuit. Critically, the QBSA module replaces the classical dot-product similarity with a quantum similarity to evaluate architectural relationships within a more expressive, high-dimensional quantum feature space. For the search objective, we employ a hardware-aware and task-agnostic target. Instead of optimizing for task-specific performance, our framework directly optimizes for fundamental physical properties of the circuit, namely, its noisy expressibility and Probability of Successful Trials (PST)\cite{WOS:000981574300135,liu2025haqgnnhardwareawarequantumkernel}, a metric used to evaluate robustness under Pauli noise. These properties serve as more direct indicators of performance and resilience on real hardware. Additionally, the framework includes a post-search optimization stage that applies a set of deterministic circuit simplification rules\cite{yan2025quantumcircuitsynthesiscompilation}, such as gate elimination, fusion, and commutation, to further refine the discovered architectures for practical hardware implementation\cite{maslov2008quantum}.

The main contributions of this work are summarized as follows:
\begin{itemize}
\item \textbf{Quantum-Native Context-Aware Architecture Search:} We introduce the QBSA-DQAS framework, which leverages a novel QBSA module to perform context-aware architecture search natively in the quantum domain. This module introduces a unified, two-stage process for quantum feature extraction: it first computes context-aware dependencies through quantum feature maps, and then enriches these representations through a powerful, position-wise non-linear quantum transformation. This integrated design enables a quantum-native approach to context-aware architecture search by ensuring all core computations are performed natively in the quantum domain, making the module better capture the relationships between quantum operations.
\item \textbf{Hardware-Aware and Task-Agnostic Multi-Objective Optimization:} We establish a hardware-aware and task-agnostic search paradigm by framing the search as a multi-objective optimization problem\cite{Ekstrom_2025}. Our framework seeks universally robust circuits by optimizing a composite objective based on fundamental physical properties, namely noisy expressibility and PST, rather than relying on task-specific performance metrics. This ensures the discovered architectures are inherently resilient to real hardware noise.
\item \textbf{Empirical Validation on Quantum Chemistry and Combinatorial Optimization:} We validate the efficacy of the QBSA-DQAS framework on two applications. For the task of computing molecular ground state energy, circuits discovered by our framework achieve high-fidelity results under realistic hardware noise. The practical utility of our framework is further demonstrated on a large-scale Wireless Sensor Network (WSN) routing problem, a representative combinatorial optimization challenge. In this application, the ansatz discovered by QBSA-DQAS yields a solution with significantly lower network energy consumption. These results confirm that our quantum-native and hardware-aware search paradigm is effective at producing robust, high-performance circuits for diverse computational challenges.
\end{itemize}

\section{Background}
\subsection{Quantum Computation}
Quantum computation is based on quantum mechanics, using qubits as basic units. Unlike classical bits, qubits exist in superposition of basis states $\ket{0}$ and $\ket{1}$. For $n$ qubits, the state space grows exponentially to dimension $2^n$, providing quantum computational advantage\cite{Montanaro2016NPJ_QCSurvey}.

Quantum circuits apply sequences of unitary gates to evolve qubit states\cite{2007QUANTUM}. Single-qubit gates include Pauli gates ($X, Y, Z$) and the Hadamard gate ($H$)\cite{davis2020towards}. Parameterized rotation gates $R_x(\theta)$, $R_y(\theta)$, $R_z(\theta)$ provide optimizable parameters for variational algorithms. Multi-qubit gates like $\mathrm{CNOT}$ and $\mathrm{CZ}$ create entanglement, a uniquely quantum correlation\cite{Horodecki2009RMP_Entanglement}. A finite gate set is universal if it enables approximation of arbitrary unitary operators to arbitrary precision\cite{kjaergaard2020superconducting}. Quantum computation concludes with measurement, which projects the quantum state onto the computational basis, yielding classical outcomes.

\subsection{Variational Quantum Algorithms}
In the current era of NISQ computing, where hardware is constrained by significant noise and limited qubit counts, VQAs have emerged as a leading computational paradigm. VQAs are hybrid quantum-classical algorithms that employ a feedback loop between a quantum processor and a classical optimizer to find solutions to complex problems. The core principle is to use a PQC, also known as an ansatz $U(\boldsymbol{\theta})$, to prepare a trial quantum state $ \ket{\psi(\boldsymbol{\theta})} = U(\boldsymbol{\theta}) \ket{0}^{\otimes n} $. The goal is to find the optimal parameters $\boldsymbol{\theta}^*$ that minimize the expectation value of a problem-specific observable, which is typically encoded in a Hamiltonian $H$. The optimization problem is formally stated as: 
\begin{align} 
\min_{\boldsymbol{\theta}} L(\boldsymbol{\theta}) = \langle\psi(\boldsymbol{\theta}) \vert H \vert \psi(\boldsymbol{\theta})\rangle 
\end{align} 

The algorithm operates iteratively. First, the quantum device prepares the state and estimates the cost function $ L(\boldsymbol{\theta}) $ through repeated measurements. This cost is then passed to a classical optimizer. For gradient-based optimizers, gradients can be estimated directly on the quantum hardware using techniques like the parameter-shift rule. For a parameter $\theta_k$, its gradient is calculated as\cite{mitarai2018quantum}:
\begin{align}
\frac{\partial L(\boldsymbol{\theta})}{\partial \theta_k} = \frac{1}{2} \left[ L(\boldsymbol{\theta} + \frac{\pi}{2}\mathbf{e}_k) - L(\boldsymbol{\theta} - \frac{\pi}{2}\mathbf{e}_k) \right]
\end{align}
where $\mathbf{e}_k$ is a unit vector along the $\theta_k$ direction. The classical optimizer then uses these gradients to propose an updated set of parameters, for instance, via gradient descent:
\begin{align}
\boldsymbol{\theta}^{(t+1)} = \boldsymbol{\theta}^{(t)} - \eta \nabla L(\boldsymbol{\theta}^{(t)})
\end{align} 
This hybrid loop continues until the cost function converges\cite{Schuld2019PRA_Gradients}. Two of the most prominent VQAs are the VQE, widely used for quantum chemistry, and the Quantum Approximate Optimization Algorithm (QAOA)\cite{farhi2014quantum}, applied to combinatorial optimization problems.

The performance of any VQA is critically dependent on the architecture of its ansatz. The design of the ansatz presents a significant challenge, as it must balance two competing properties: expressibility and trainability. Expressibility refers to the circuit's ability to generate a sufficiently rich set of states to represent the problem's solution. However, highly expressive ansatzes, particularly those that are deep and unstructured, often suffer from the barren plateau phenomenon. In this phenomenon, cost function gradients vanish exponentially with the number of qubits, rendering the optimization intractable. Common manual design strategies, such as the problem-inspired Unitary Coupled Cluster with Single and Double Excitations (UCCSD) and Hardware-Efficient Ansatzes (HEAs)\cite{mcclean2016theory}, represent different trade-offs in this balance, but often struggle to achieve optimal performance. The inherent difficulty in this manual design process motivates the development of automated methods to discover more effective circuit architectures.

\subsection{The Self-Attention Mechanism in Classical Machine Learning}
The field of machine learning has seen significant advancements in processing sequential data, largely driven by the development of attention mechanisms. Traditional models like Recurrent Neural Networks (RNNs) process sequences element by element, often struggling to capture long-range dependencies due to the vanishing gradient problem. The self-attention mechanism, a key component of the Transformer architecture, provides a powerful solution by calculating the dependency between any two elements in the sequence in parallel through matrix operations, thus overcoming the sequential processing limitations of RNNs. It allows a model to weigh the importance of different elements within a single input sequence to compute a contextualized representation for each element, regardless of their distance from one another.

The mechanism operates on a set of input vectors, which are first projected into three distinct representations: a Query matrix ($Q$), a Key matrix ($K$), and a Value matrix ($V$), through learned linear transformations. The attention score for each element is computed by taking the dot product of its Query vector with the Key vectors of all elements in the sequence. These scores are then scaled, typically by the square root of the key dimension $d_k$, to maintain stable gradients. A softmax function is applied to the scaled scores to obtain normalized attention weights. The final output for each element is a weighted sum of all Value vectors, where the weights are the computed attention probabilities. The entire operation can be expressed as\cite{vaswani2017attention}:
\begin{align}
\text{Attention}(Q, K, V) = \text{softmax}\left(\frac{QK^T}{\sqrt{d_k}}\right)V
\end{align}

A crucial enhancement to this mechanism is Multi-Head Attention. Instead of performing a single attention function, this approach linearly projects the queries, keys, and values multiple times with different, learned projections. Attention is then computed in parallel for each of these projected versions. This allows the model to jointly attend to information from different representation subspaces at different positions. The outputs of the parallel heads are then concatenated and linearly projected again to produce the final result. Since the self-attention mechanism itself is permutation-invariant, positional information is typically added to the input embeddings using fixed or learned positional encodings to inform the model about the order of the sequence. The success of the Transformer architecture across numerous domains has established self-attention as a fundamental and highly effective tool for learning complex relationships within structured data\cite{Khan2023TPAMI_TransformerSurvey}.

\begin{figure*}[t]
    \centering
    \includegraphics[width=1.0\textwidth]{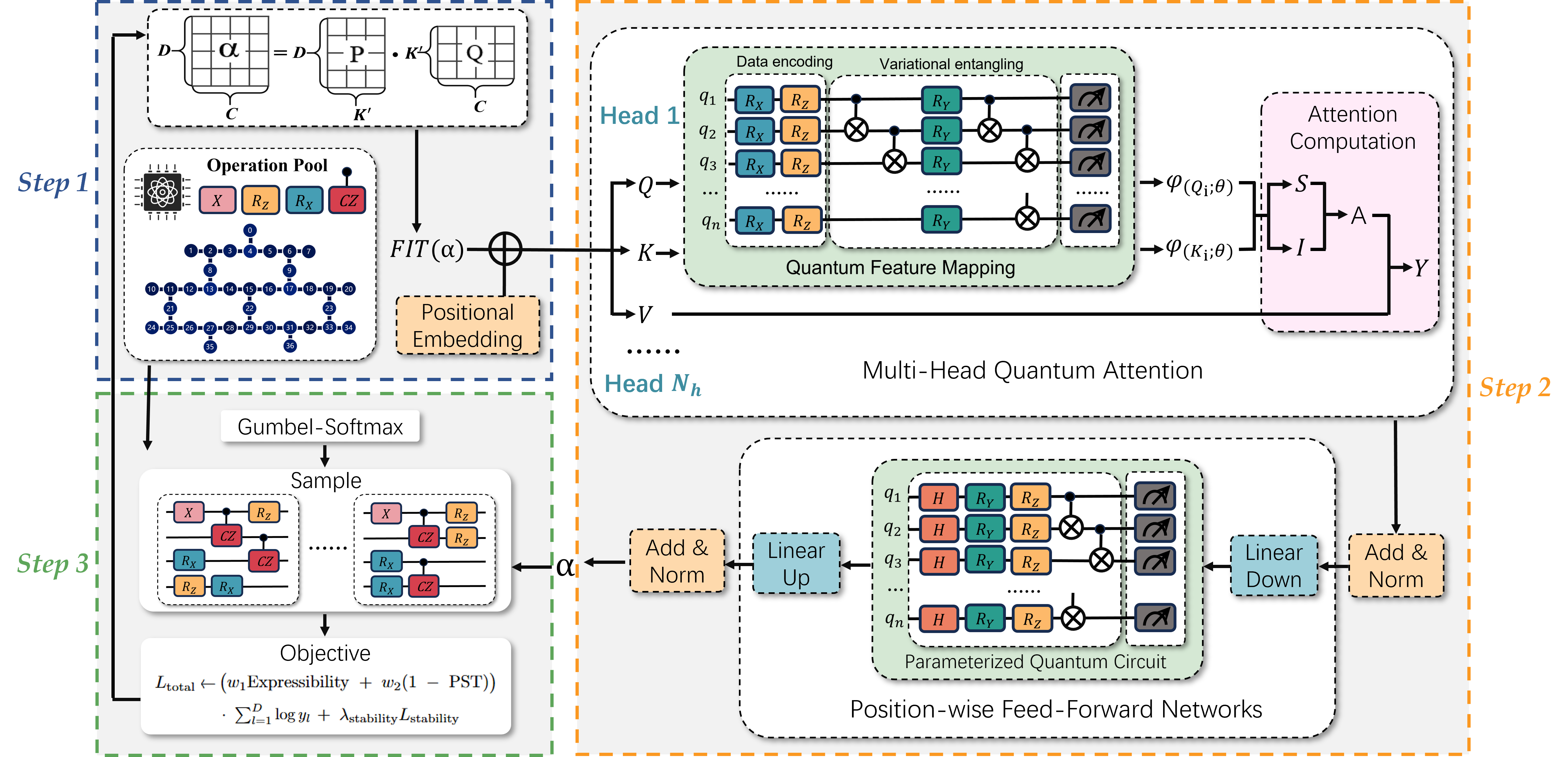} 
    \caption{An overview of the QBSA-DQAS architecture search algorithm.} \label{fig1}
\end{figure*}
\section{Method}

In this section, we present the technical details of our proposed QBSA-DQAS framework. A schematic overview of the entire framework is presented in Figure \ref{fig1}, which consists of four main steps:

\textbf{\textit{Step 1:} Hardware-Aware Architecture Space Construction.} We define a search space tailored to the target quantum hardware, encoding the architectural choices into a parameter matrix $\alpha$ that respects the device's topology and native gate set.

\textbf{\textit{Step 2:} Quantum-Based Self-Attention Module.} We introduce the QBSA module, designed to perform context-aware feature extraction natively in the quantum domain. The module operates via a unified, two-stage process. First, to capture the contextual dependencies among architectural choices, it computes a similarity matrix using quantum feature maps. This similarity computation evaluates the relationships between architectural elements by mapping them into a high-dimensional Hilbert space, allowing it to capture complex quantum interactions. Second, the module processes these contextual representations through a position-wise quantum non-linear transformation to further enrich the features. This integrated design ensures that the entire feature extraction process is performed natively in the quantum domain.

\textbf{\textit{Step 3:} Hardware-Aware Differentiable Architecture Search.} We frame the search as a multi-objective optimization problem, guided by hardware-aware and task-agnostic metrics, namely noisy expressibility and PST. This approach seeks to discover architectures that are inherently robust and high-performing. To navigate the search space efficiently, we employ a differentiable sampling strategy based on the Gumbel-Softmax technique\cite{jang2017categoricalreparameterizationgumbelsoftmax} with adaptive temperature annealing.

\textbf{\textit{Step 4:} Post-Search Circuit Optimization.} A deterministic, rule-based routine is applied to the discovered circuits to reduce gate count and depth by eliminating redundancies, fusing compatible gates, and reordering operations based on commutation rules.
\begin{figure*}[t]
    \centering
    \includegraphics[width=0.92\textwidth]{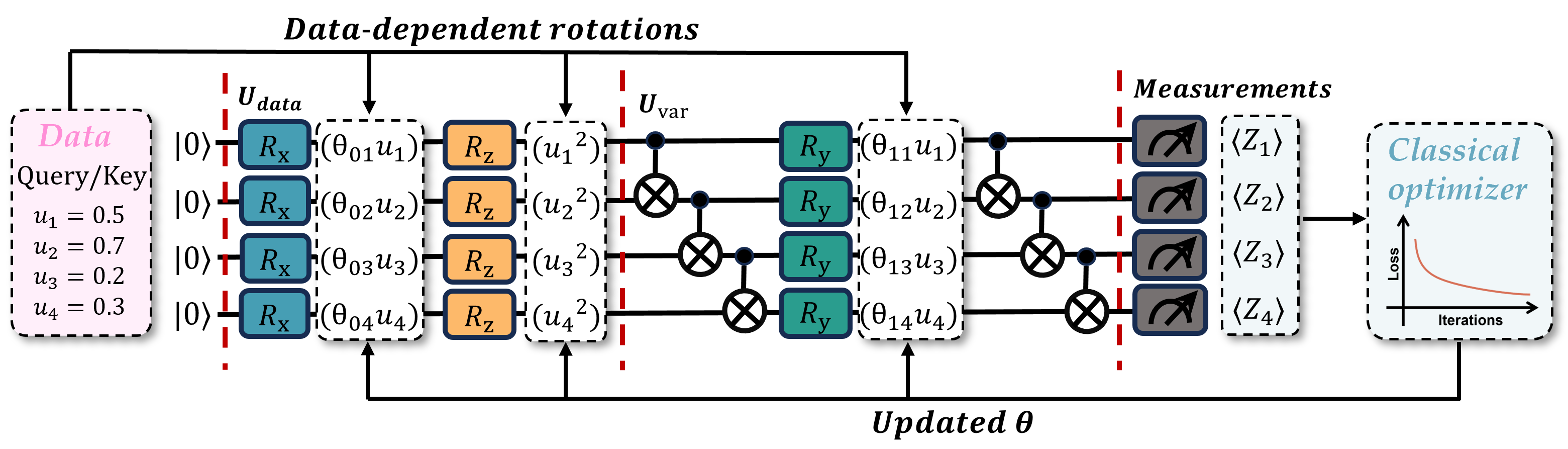} 
    \caption{Quantum feature map circuit for attention.}\label{circuit1}
\end{figure*}
\subsection{Hardware-Aware Architecture Space Construction}

The foundation of our QBSA-DQAS framework is a search space constructed to be aware of the target hardware's physical constraints. This stage constructs the operation pool based on device topology and native gate capabilities, followed by defining the architectural parameter matrix $\alpha$. This hardware-centric design ensures that all discovered circuits can be directly implemented without requiring costly gate decomposition or SWAP operations, thus maximizing fidelity and performance.

The architectural parameter matrix $\alpha \in \mathbb{R}^{D \times C}$ serves as the core representation of our differentiable search space, where $D$ represents the maximum circuit depth and $C$ corresponds to the size of the operation pool. Each element $\alpha_{d,c}$ encodes the unnormalized log-probability for selecting operation $c$ at circuit depth $d$. The matrix provides a continuous, differentiable representation that enables gradient-based optimization with its dimensions and content defined by the hardware's constraints.

To address the computational challenges and mitigate the risk of overfitting in the large search space, we employ a low-rank matrix factorization for parameterizing the architectural parameters.
\begin{align}
\alpha = PQ
\end{align}
where $P \in \mathbb{R}^{D \times 1 \times K'}$ and $Q \in \mathbb{R}^{D \times K' \times C}$, with $K'$ representing the factorization rank. The choice of $K'$ involves a trade-off: larger values increase the model's capacity to capture complex patterns in the architecture space, while smaller values provide stronger dimensionality reduction and a more constrained search, which can be beneficial for avoiding overfitting.

Preprocessing applies Feature Interaction Transformation (FIT):
\begin{align}
\alpha' = (\alpha \alpha^T) \alpha
\end{align}
This transformation creates higher-order interaction terms that capture relationships between circuit depth positions, enriching features with global architectural context.
Second, sinusoidal positional encoding injects circuit depth information into the parameter matrix (formulation in Appendix A). The final preprocessed matrix $\alpha_{\mathrm{in}} = \alpha' + \text{PE}$ is then passed to the QBSA module.

\subsection{Quantum-Based Self-Attention Mechanism}
\begin{figure*}[t]
    \centering
    \includegraphics[width=0.76\textwidth]{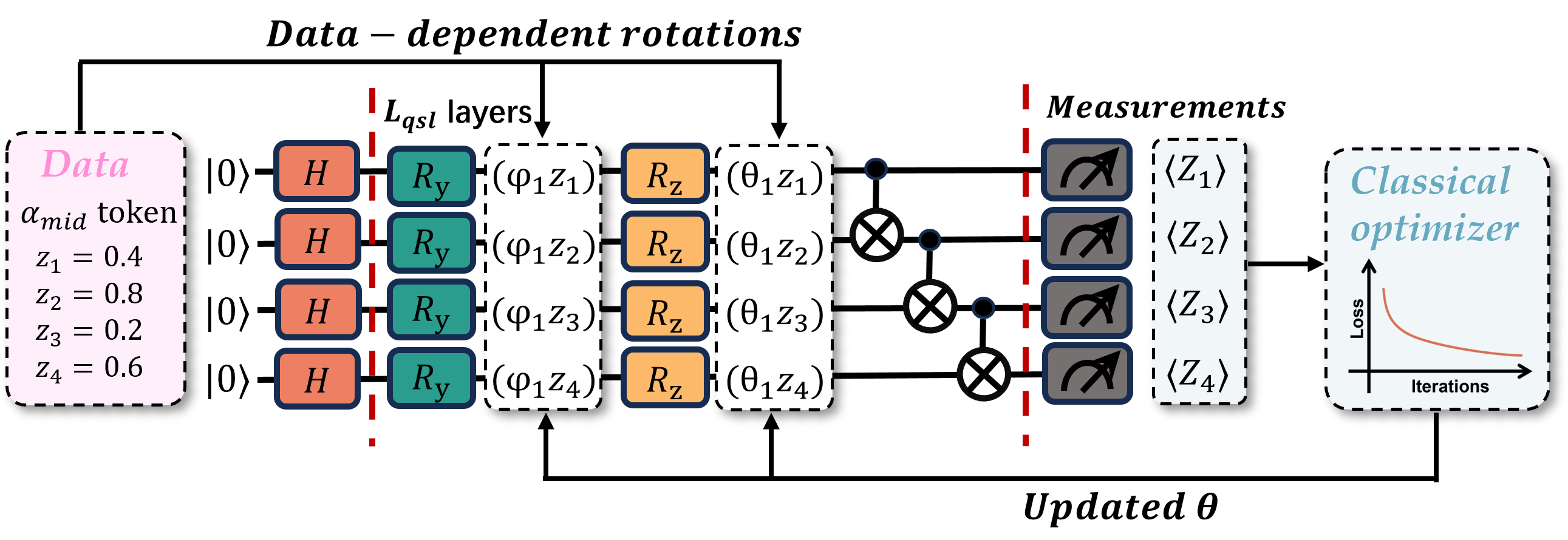} 
    \caption{Position-wise feed-forward quantum circuit.}\label{circuit2}
\end{figure*}
The QBSA module is a unified quantum information processing unit that transforms $\alpha_{\mathrm{in}}$ into $\alpha_{\mathrm{out}}$ through two tightly coupled stages. First, it computes similarity scores combining quantum feature similarity with a phase-controlled interference term, producing intermediate representation $\alpha_{\mathrm{mid}}$. Second, it applies position-wise quantum transformation leveraging a compact data-reuploading ansatz\cite{P_rez_Salinas_2020}. Non-linearity is introduced through expectation value measurements. This design ensures both stages operate natively in the quantum domain.

Input $\alpha_{\mathrm{in}}$ is projected into query, key, and value representations via learnable transformations (formulation in Appendix A). Quantum feature mapping transforms query and key vectors into quantum Hilbert space. For each vector $u$, a parameterized quantum circuit prepares quantum state $\lvert \psi(u;\theta)\rangle$, through a two-stage architecture comprising data encoding and variational entangling unitaries. For an $n$-qubit system, this is formulated as:
\begin{align}
U_{\mathrm{data}}(u;\theta_0) = \left(\bigotimes_{i=1}^{n} R_x(\theta_{0,i} u_i)\right) \left(\bigotimes_{i=1}^{n} R_z(u_i^2)\right)
\end{align}
where the learnable scaling parameters $\theta_{0,i}$ in the $R_x$ rotations enable adaptive weighting of input features, while the quadratic terms in the $R_z$ rotations introduce crucial nonlinearity.
\begin{align}
U_{\mathrm{var}}(u;\theta_1)
&= \left( \prod_{i=1}^{n-1} \mathrm{CNOT}_{i,i+1} \right)
   \left( \bigotimes_{i=1}^{n} R_y(\theta_{1,i} u_i) \right) \notag \\
&\quad \left( \prod_{i=1}^{n-1} \mathrm{CNOT}_{i,i+1} \right) \label{eq:Uvar}
\end{align}
The quantum state $\ket{\psi(u;\ \theta)}=U_{\text{var}}(u;\ \theta_{1})U_{\text{data}}(u;\ \theta _{0})\ket{0}^{\otimes n}$ is prepared through the circuit architecture shown in Figure \ref{circuit1}. From the quantum state, we extract feature vectors $\phi(u;\theta) \in \mathbb{R}^n$ by measuring the Pauli-Z expectation value on each qubit. This measurement process is differentiable via the parameter-shift rule, enabling end-to-end gradient-based optimization. The quantum feature map implicitly projects inputs into an exponentially large ($2^n$-dimensional) Hilbert space while using only a linear number of trainable parameters.

Similarity is computed through quantum-derived features and interference. The quantum feature similarity is
\begin{align}
S_{ij} = \phi(Q_i;\theta)^\top \phi(K_j;\theta)
\end{align}
whose properties are learned during training, unlike a fixed dot-product.
The quantum interference term is:
\begin{align}
I_{ij} = N_h \cdot \|\phi(Q_i;\theta)\|_2 \cdot \|\phi(K_j;\theta)\|_2 \cdot \cos(\varphi^{(r)})
\end{align}
where $N_h$ is the number of heads and $\varphi^{(r)}$ is a learnable phase. This mechanism allows the model to learn constructive or destructive interference patterns, capturing complex dependencies. Combined logits $\Xi_{ij} = S_{ij} + I_{ij}$ are transformed into attention weights via scaled softmax normalization, which then aggregate value vectors to produce head outputs:
\begin{align}
A_{ij} = \frac{\exp(\Xi_{ij} / (\sqrt{d_h} \cdot \tau))}{\sum_{k=1}^{C} \exp(\Xi_{ik} / (\sqrt{d_h} \cdot \tau))}
\end{align}
The mechanism operates within a multi-head framework, where each head maintains independent projection matrices, quantum circuit parameters, and interference phases. Multi-head outputs are concatenated, projected, and integrated with the input via residual connection and layer normalization to produce intermediate representation $\alpha_{\mathrm{mid}}$ (formulations in Appendix A).

The second stage applies a position-wise transformation $T$ independently to each token in $\alpha_{\mathrm{mid}}$. Each token is first linearly projected to $z \in \mathbb{R}^{n_{\mathrm{qubits}}}$, then processed by a compact parameterized quantum circuit of depth $L_{qsl}$. The circuit employs Hadamard initialization followed by layers of data-dependent rotations and $\mathrm{CNOT}$ entangling gates:
\begin{align}
U_{l}(z) = \left(\prod_{i=1}^{n-1} \mathrm{CNOT}_{i,i+1}\right)\left(\bigotimes_{i=1}^{n} R_{Y}(\theta_{l,i} z_i) R_{Z}(\phi_{l,i} z_i)\right)
\end{align}
After the final layer, a feature vector $s$ is extracted by measuring the Pauli-Z expectation value of each qubit. This layer structure is illustrated in Figure \ref{circuit2}. A final linear projection restores the original dimension. The overall module output is produced by integrating this transformation with residual connection and layer normalization:
\begin{align}
\alpha_{\mathrm{out}} = \mathrm{LayerNorm}\left(\alpha_{\mathrm{mid}} + \mathrm{Dropout}(T(\alpha_{\mathrm{mid}}))\right)
\end{align}

\subsection{Hardware-Aware Differentiable Architecture Search}

This section details our differentiable search framework, designed to identify quantum architectures that are both performant and robust against hardware noise by optimizing a hardware-aware objective function via a differentiable sampling technique.

Gumbel-Softmax sampling enables differentiable exploration of discrete gate choices. For each gate position $d$, the probability of selecting gate $k$ is computed as:
\begin{align}
\tilde{h}_d^{(k)} = \frac{\exp\left(\frac{(\alpha_{\mathrm{out}})_d^{(k)} + G_{k}}{T}\right)}{\sum_{l=1}^C \exp\left(\frac{(\alpha_{\mathrm{out}})_d^{(l)} + G_{l}}{T}\right)}
\end{align}
where $G_k = -\log(-\log(U_k))$ represents Gumbel noise with $U_k \sim \text{Uniform}(0,1)$, and $T$ is a temperature parameter that is gradually decreased during training via an annealing schedule. This process encourages exploration by initially allowing for soft, probabilistic selections before converging to hard, discrete choices, while maintaining gradient flow via the straight-through estimator\cite{Bengio2013EstimatingOP}.

Our optimization is guided by a hardware-aware objective function that combines two complementary metrics: expressibility and a fidelity proxy. Expressibility quantifies state space coverage via Kullback-Leibler (KL) divergence:
\begin{align}
\text{Expressibility} &= D_{\text{KL}}(P_{\text{circuit}} \parallel P_{\text{Haar}}) \notag\\
&= \sum_{F} P_{\text{circuit}}(F) \log_{2} \left( \frac{P_{\text{circuit}}(F)}{P_{\text{Haar}}(F)} \right)
\label{exp}
\end{align}
where lower values indicate higher expressibility. To evaluate the circuit's resilience to noise, we adopt the PST as a proxy for fidelity. Instead of costly direct fidelity estimation, we concatenate each circuit $U$ with its inverse $U^\dagger$ and apply the combined circuit to the initial state $\lvert 0\rangle^{\otimes n}$. The PST is then defined as the proportion of measurements that yield the initial state:
\begin{align}
\text{PST} = \frac{T_{\text{initial}}}{T_{\text{total}}}
\end{align}
where $T_{\text{initial}}$ is the number of trials with an output identical to the initial state, and $T_{\text{total}}$ is the total number of trials. This metric directly assesses the circuit's computational stability under noise, with higher values indicating greater robustness. The search is guided by a composite loss function that encourages the selection of architectures with lower cost, while a regularization term ensures a stable training trajectory. The cost, $C_k$, for a sampled circuit $k$ is defined as a weighted sum of its performance metrics:
\begin{align}
C_k = w_1 \cdot \text{Expressibility} + w_2 (1 - \text{PST})
\end{align}
where $w_1$ and $w_2$ are balancing coefficients.

To ensure the search converges smoothly, we introduce a stability penalty, $L_{\text{stability}}$, which discourages erratic changes in the encoder's output, $F_t$, across consecutive training steps. It is defined using the $L_\infty$ norm:
\begin{align}
L_{\text{stability}} = \max_{i,j} \left\vert F_t(a_{ij}^{\text{trans}}) - F_{t-1}(a_{ij}^{\text{trans}}) \right\vert
\end{align}

The complete loss function, $L_{\text{total}}$, is defined as the batch-averaged sum of the cost-weighted log-probabilities and the stability penalty:
\begin{align}
L_{\text{total}} = \frac{1}{B} \left( \sum_{k=1}^{B} \left( C_k \sum_{i} \log(p_{k,i}) \right) + \lambda_{\text{stability}} \cdot L_{\text{stability}} \right)
\end{align}
where $B$ is the batch size, $p_{k,i}$ is the probability of selecting operation $i$ in circuit $k$, and $\lambda_{\text{stability}}$ controls the strength of the stability regularization. By minimizing this loss, the algorithm learns to increase the selection probability of architectures with lower costs, effectively guiding the search towards structures that are both performant and reachable through a stable optimization process. Our complete search algorithm integrates the hardware-aware search space, the QBSA module, and the hardware-aware objective into a complete end-to-end procedure, which is formally summarized in Algorithm \ref{alg:QBSA_dqas}.

\begin{figure}[ht!]
\refstepcounter{algorithm}
\hrule
\vspace{0.2em}
\noindent\textbf{Algorithm 1: The QBSA-DQAS Algorithm}
\label{alg:QBSA_dqas}
\vspace{0.2em}
\hrule
\vspace{0.3em}

\begin{algorithmic}[1]
    \Require Operation pool $\Omega$, hardware noise model $\mathcal{N}$, training steps $T$, temperature schedule $\{\tau_t\}$
    \Ensure Optimized circuit architecture $C^*$
    \Function{QBSA-DQAS}{$\Omega, \mathcal{N}, T, \{\tau_t\}$}
      \State Initialize low-rank factors $P, Q$ and attention parameters
      \For{$t = 1$ \textbf{to} $T$}
        \State $\alpha \gets P Q$ \Comment{Construct architecture logits}
        \State $\alpha_{\mathrm{in}} \gets \mathrm{FIT}(\alpha) + \mathrm{PE}$ \Comment{Preprocess with FIT transform and positional encoding}
        
        \ForAll{head $h \in \{1, \dots, N_h\}$} \Comment{Multi-head quantum self-attention}
            \State Project $\alpha_{\mathrm{in}}$ to $Q^{(h)}, K^{(h)}, V^{(h)}$
            \State Build quantum features $\phi(\cdot)$ via PQC for rows of $Q^{(h)}, K^{(h)}$
            \State Compute attention logits $\Xi^{(h)}$ from quantum feature similarity $S^{(h)}$ and interference $I^{(h)}$
            \State Normalize attention logits to obtain weights $A^{(h)}$, aggregate values to $Y^{(h)}$
        \EndFor
        
        \State $Y_{\mathrm{cat}} \gets \mathrm{Concat}(Y^{(1)},\dots,Y^{(N_h)})$
        \State $\alpha_{\mathrm{mid}} \gets \mathrm{LayerNorm}\big(\alpha_{\mathrm{in}} + Y_{\mathrm{cat}} W_o\big)$ \Comment{Integrate attention output via residual fusion}
        
        \State $output \gets \mathrm{Linear}_{\mathrm{up}}(\mathrm{PQC}_{\mathrm{Zexp}}(\mathrm{Linear}_{\mathrm{down}}(\alpha_{\mathrm{mid}})))$ \Comment{Apply position-wise PQC transformation}
        \State $\alpha_{\mathrm{out}} \gets \mathrm{LayerNorm}\!\big(\alpha_{\mathrm{mid}} + \mathrm{Dropout}(output)\big)$ \Comment{Integrate transformation via residual fusion}
        
        \State $y \sim \mathrm{GumbelSoftmax}(\alpha_{\mathrm{out}}, \tau_t)$ \Comment{Reparameterized sampling}
        
        \State Estimate $\mathrm{Expressibility}$ and $\mathrm{PST}$ for circuit $y$ under $\mathcal{N}$ \Comment{Hardware-aware evaluation}
        \State Compute $L_{\mathrm{stability}}$ from consecutive encoder outputs \Comment{Stability regularization}
        
        \State $L_{\mathrm{total}} \gets \big(w_1 \mathrm{Expressibility} + w_2(1-\mathrm{PST})\big)\cdot \sum_{l=1}^{D}\log \mathrm{softmax}(\alpha_{\mathrm{out}})[l, y_l] + \lambda_{\mathrm{stability}}L_{\mathrm{stability}}$
        
        \State Update $\{P,Q\}$ and attention parameters via backpropagating $\nabla L_{\mathrm{total}}$
      \EndFor
      \State \Return $C^*$
    \EndFunction
\end{algorithmic}

\vspace{0.2em}
\hrule
\end{figure}

\begin{figure*}[t]
    \centering
    \includegraphics[width=0.98\textwidth]{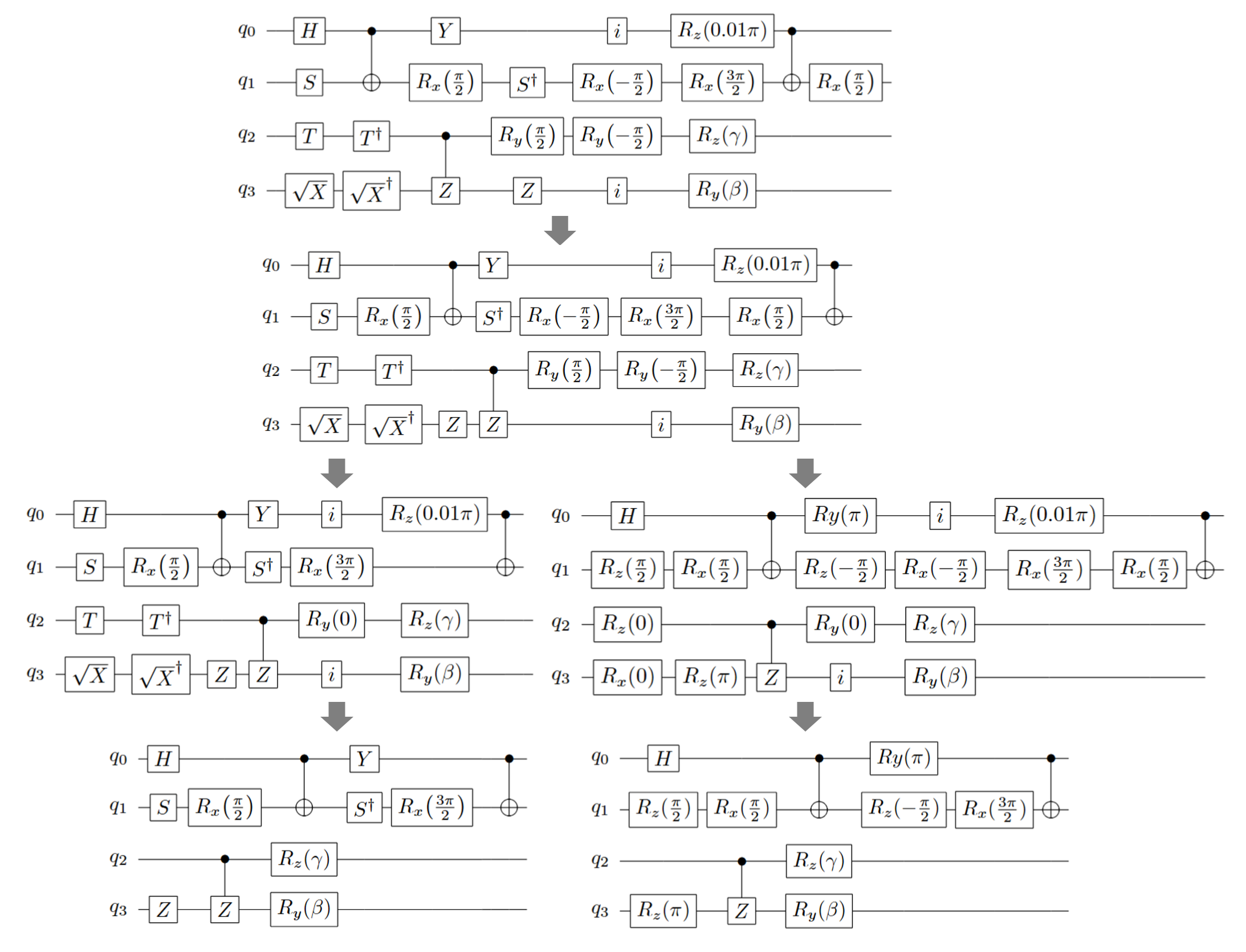} 
    \caption{The post-search optimization cascade.} \label{fig2}
\end{figure*}
\subsection{Post-Search Circuit Optimization}

Following architecture search, a post-processing optimization routine refines the discovered circuits for execution on near-term quantum hardware. This is critical because the cumulative effect of hardware imperfections, such as gate errors, thermal relaxation, and readout errors, degrades computational accuracy. Our optimization systematically reduces gate count and circuit depth. A lower gate count reduces cumulative error probability, while a shallower depth shortens execution time and limits decoherence. We employ a hierarchical cascade of strategies applied iteratively until maximal compression.

The optimization begins with gate reordering via commutation rules. This strategy pushes single-qubit gates through two-qubit gates based on their commutation properties. While not reducing gates directly, reordering creates new adjacencies by grouping previously separated compatible gates, enabling subsequent fusion and elimination. The simplification stage combines gate fusion and elimination. Gate fusion merges adjacent rotation gates of the same type by summing their rotation angles to reduce single-qubit gate count. Fusion operates in two modes. The conservative mode merges only existing rotation gates, while the aggressive mode first converts non-parametric gates like $X$ or $S$ into their rotational equivalents such as $R_x(\pi)$ or $R_z(\pi/2)$ to create additional fusion opportunities. Gate elimination then removes redundant operations. This includes canceling adjacent inverse gate pairs such as $S$ and $S^\dagger$, removing consecutive self-inverting two-qubit gates like adjacent $\mathrm{CNOT}$s, eliminating identity gates, and pruning rotation gates with negligibly small angles. The entire cascade of reordering, fusion, and elimination is repeated until no further reductions are possible. An illustrative example of this optimization cascade is provided in Figure \ref{fig2}.

This optimization ensures the final circuits are maximally compressed, reducing operational complexity for reliable execution on noisy hardware.

\section{Evaluation}

\subsection{VQE for Molecular Ground State Energy}

\subsubsection{Task Description and Setup}
This task aims to compute the ground state energy of three molecules, $H_2$, $LiH$, and $BeH_2$, using the VQE algorithm. These molecules correspond to quantum simulation systems requiring 4, 6, and 8 qubits, respectively. Performance is evaluated using the absolute energy error $\Delta E$, defined as $\Delta E = \vert E_{VQE} - E_{FCI} \vert$,  where $E_{VQE}$ denotes the energy computed via VQE, and $E_{FCI}$ is the exact energy from a classical full configuration interaction (FCI) solver. An error below 0.1 Hartree is considered to indicate a high-quality result.

\begin{figure*}[t]
    \centering
    \includegraphics[width=0.92\textwidth]{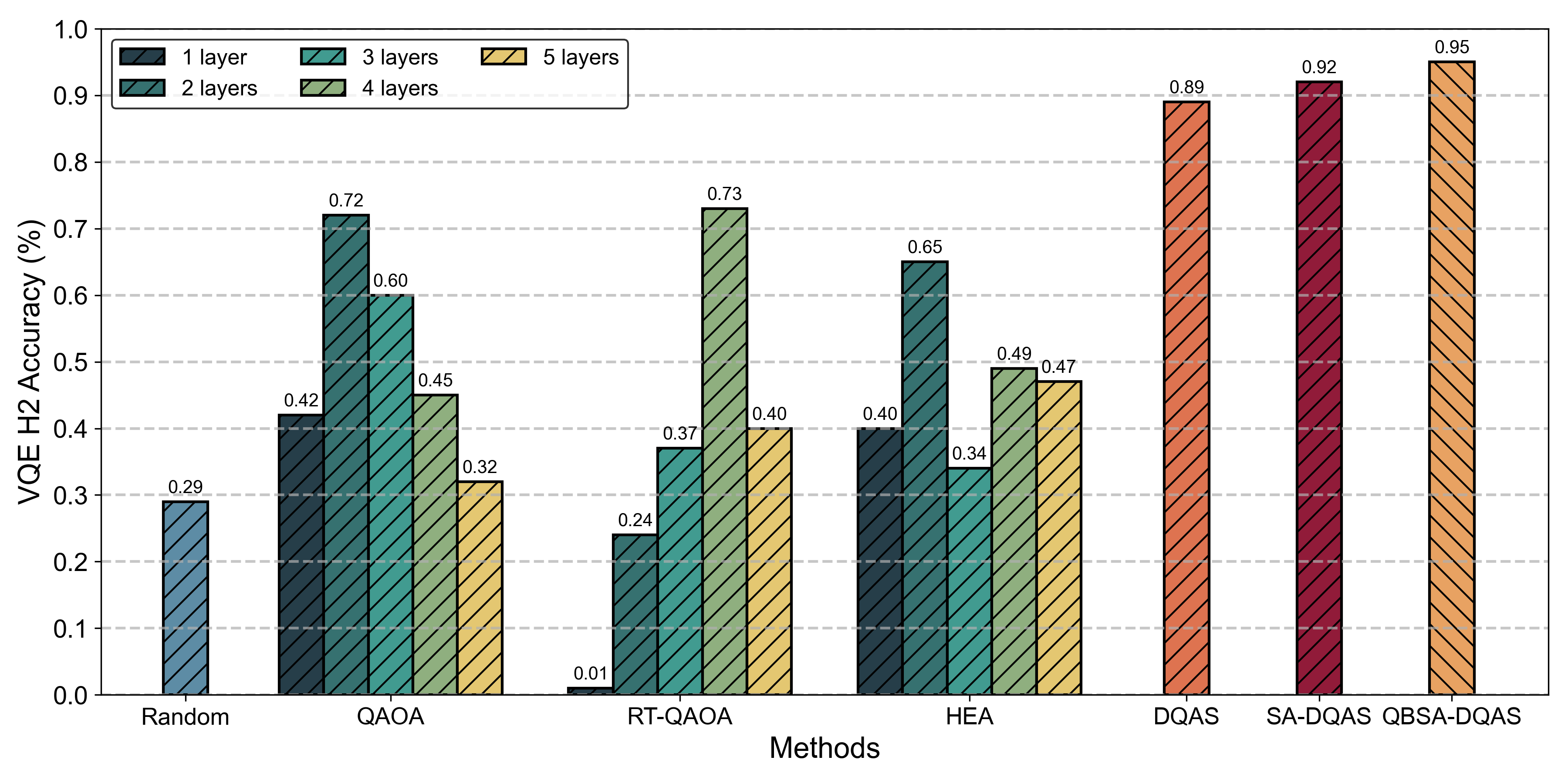} 
    \caption{VQE accuracy comparison for the $H_2$ molecule in a noiseless environment.} 
    \label{fig3}
\end{figure*}
All experiments were conducted using a unified software stack: PyTorch was used for the architecture search framework, and PennyLane was used for quantum circuit simulations. The operation pool for all circuit constructions consists of a universal gate set chosen for its broad compatibility with near-term quantum hardware, including the X gate, parameterized rotations $R_x(\theta)$ and $R_z(\theta)$, and the entangling $\mathrm{CZ}$ gate. The VQE parameter optimization for each circuit was performed for a maximum of 300 iterations, utilizing the Adam optimizer with a learning rate of 0.1. To simulate realistic noisy environments, we used hardware noise models based on five real IBM quantum devices: IBM Fez, IBM Kingston, IBM Marrakesh, IBM Pittsburgh, and IBM Torino.

\subsubsection{Baseline Methods}
To contextualize the performance of our proposed framework for this task, we established a set of baseline methods for comparison, which are grouped into two categories.

The Differentiable Architecture Search Baselines include:
\begin{itemize}
    \item DQAS (Differentiable Quantum Architecture Search): A foundational differentiable search framework that employs Gumbel-Softmax reparameterization to relax discrete gate selection into continuous optimization. It parameterizes the architecture space through learnable weight matrices and enables gradient-based optimization for efficient exploration of circuit structures. 
    \item SA-DQAS (Self-Attention DQAS)\cite{sun2024sa}: This method extends DQAS by incorporating a classical Transformer-based self-attention encoder to capture contextual dependencies among quantum gates. While the self-attention mechanism enables modeling of long-range gate interactions, it operates entirely in the classical domain using standard dot-product attention.
\end{itemize}

The Fixed-Ansatz Baselines include:
\begin{itemize}
    \item Random: This approach generates circuits by randomly selecting gates and their positions from a predefined operation pool.
    \item QAOA: A problem-inspired ansatz that constructs circuits by alternating between problem Hamiltonian evolution operators (typically $\mathrm{CNOT}-R_z-\mathrm{CNOT}$ and $R_z$ gates) and mixer Hamiltonian operators ($R_x$ gates). Each layer introduces two trainable parameters. Its rigid layer structure exhibits theoretical guarantees for certain optimization problems.
    \item RT-QAOA (Rapidly trainable and shallow-compiled QAOA)\cite{rt-qaoa}: This ansatz enhances standard QAOA through two key modifications. It employs a compact $R_z-\mathrm{CNOT}$ structure in the first layer to reduce circuit depth, and compresses the parameter search space to $[0, \pi]$ for the initial $p-1$ layers while maintaining $[0, 2\pi]$ for the final layer. These modifications can improve trainability and reduce compilation overhead on near-term hardware.
    \item HEA: This ansatz is constructed from repeating blocks of parameterized single-qubit rotations (typically $R_y$ and $R_z$ gates) followed by a fixed pattern of two-qubit entangling gates (such as $\mathrm{CNOT}$ or $\mathrm{CZ}$ gates). The structure is designed to align with native gate sets of quantum hardware, minimizing compilation depth. 

\end{itemize}

\subsubsection{Results and Analysis}
To evaluate the effectiveness of our proposed QBSA-DQAS framework, we conducted four experiments focusing on distinct performance aspects: architecture search in noiseless environments, the impact of the hardware-aware objective function, robustness across different hardware noise models, and the effect of post-search circuit optimization.

\textbf{Performance in Noiseless Environments.} We evaluated the performance of QBSA-DQAS against all baseline methods in a noiseless simulation environment for the $H_2$ molecule. The results, summarized in Figure \ref{fig3}, shows the superiority of our approach. 

Our QBSA-DQAS framework achieves the highest accuracy of 0.95, outperforming both the standard DQAS (0.89) and the classical attention-based SA-DQAS (0.92). In contrast, the fixed-ansatz baselines failed to consistently identify effective circuit structures, frequently converging to solutions with significantly higher energy errors. Furthermore, their performance is highly sensitive to the number of layers. For instance, the accuracy of QAOA peaks at 2 layers and then declines, which highlights the difficulty of manually determining an optimal depth for such ansatzes. This analysis demonstrates the effectiveness of our QBSA-DQAS framework. Our QBSA mechanism captures the complex interactions between quantum gates more effectively, enabling a more efficient search for highly expressive and problem-specific circuit architectures compared to other search methods.

\begin{figure}[t]
    \centering
    \includegraphics[width=0.5\textwidth]{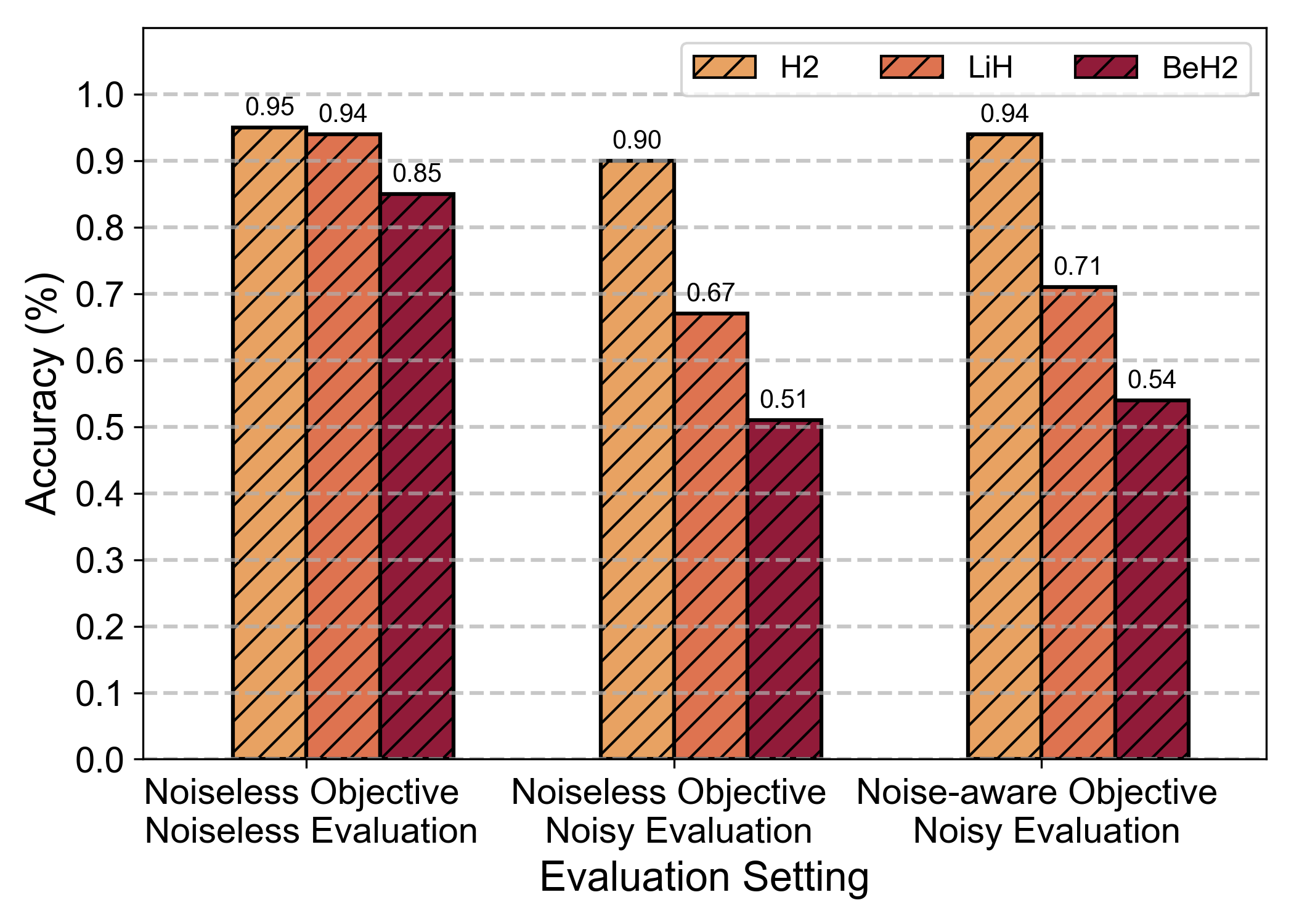} 
    \caption{VQE accuracy for different objectives across ideal and noisy simulation environments.} 
    \label{hardware_ablation}
\end{figure}

\textbf{Ablation Study on the Noise-Aware Objective.} To validate the contribution of our noise-aware objective function, we conducted an ablation study, with the results presented in Figure \ref{hardware_ablation}. We compared circuits optimized with a standard noiseless objective against those optimized with our hardware-aware objective, evaluating them in both ideal and noisy simulation environments. 

The results clearly demonstrate the benefit of the hardware-aware approach. While circuits optimized for a noiseless setting achieve high accuracy in ideal conditions, their performance degrades substantially when subjected to noise, with the accuracy of $BeH_2$ plummeting from 0.85 to 0.51. This highlights the vulnerability of circuits optimized without considering noise. In contrast, incorporating the hardware-aware objective during the search enables our framework to discover circuits with greater resilience. Under the same noisy evaluation, these circuits achieve improved accuracy across all molecules, with the accuracy of $LiH$ rising from 0.67 to 0.71. This analysis confirms that our hardware-aware objective is essential for guiding the search towards architectures robust against hardware noise, making them more suitable for deployment on real quantum devices.

\begin{figure}[t]
    \centering
    \includegraphics[width=0.5\textwidth]{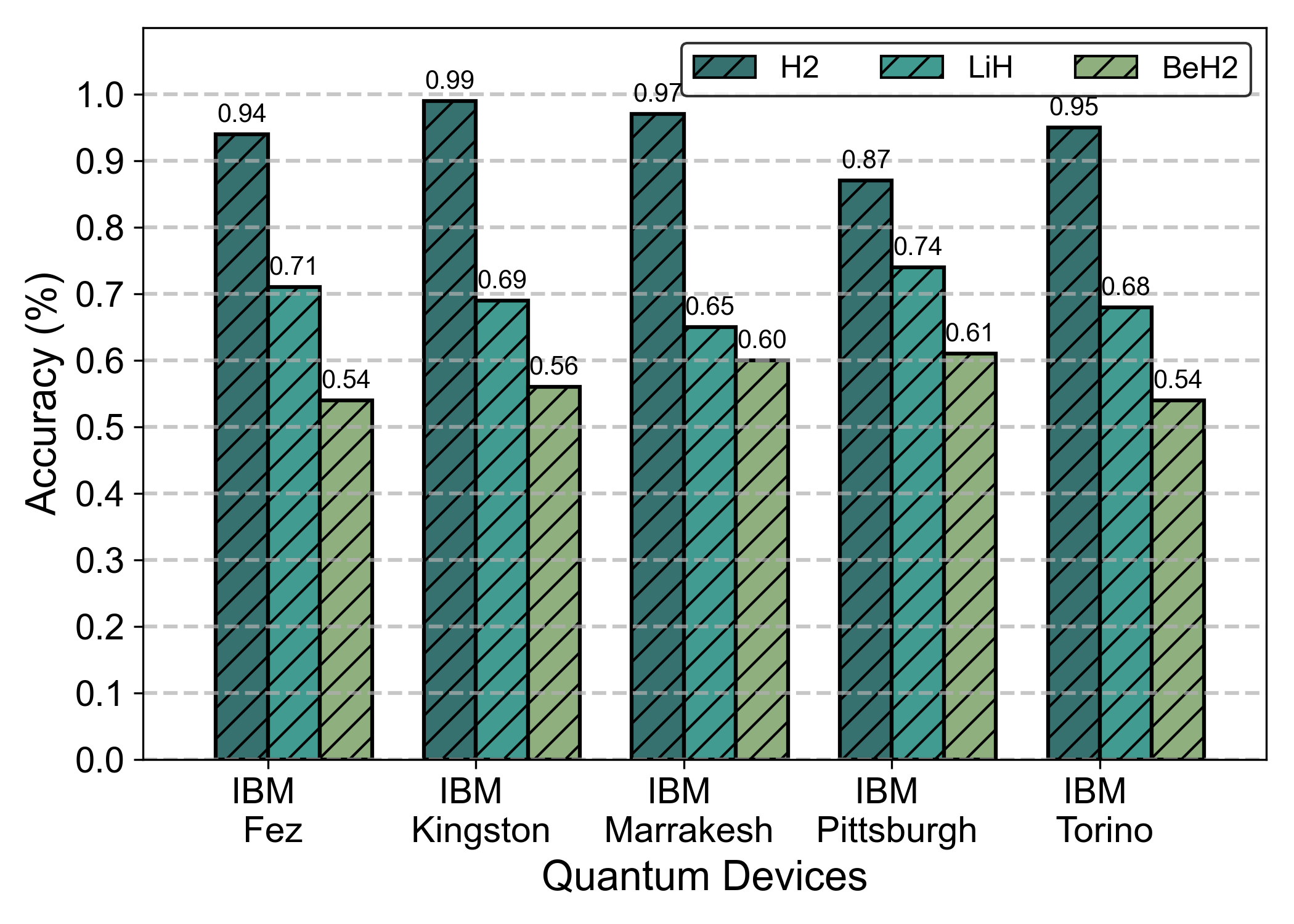} 
    \caption{VQE accuracy for different molecules across various IBM hardware noise models.} 
    \label{fig4}
\end{figure}

\textbf{Framework Robustness Across Hardware Noise Models.} This experiment assessed the robustness of our QBSA-DQAS framework across a diverse set of realistic hardware noise profiles. We independently conducted the entire architecture search and optimization process for three molecules, including $H_2$, $LiH$, and $BeH_2$, under five distinct hardware noise models from IBM Quantum devices, as presented in Figure \ref{fig4}. 

The results show that our framework successfully discovered high-performance circuits for all molecules in each of the five noisy environments. For the $H_2$ molecule, the accuracy remained consistently high, ranging from 0.87 under the IBM Pittsburgh noise model to 0.99 under the IBM Kingston noise model. The performance degradation observed for larger molecules is physically interpretable, as their greater intrinsic physical complexity demands simulations with more qubits and deeper circuits, which inherently amplifies the cumulative impact of hardware noise. This demonstrates that the effectiveness of QBSA-DQAS is not limited to a single specific hardware noise condition. Instead, the framework consistently discovers noise-resilient solutions across diverse hardware conditions, demonstrating its practical utility for NISQ processors.

\textbf{Impact of Post-Search Circuit Optimization.}
This experiment evaluated the effectiveness of our post-search optimization routine in simplifying the discovered circuit architectures and enhancing their performance under realistic noise conditions. The process was applied to the circuits discovered for the $H_2$, $LiH$, and $BeH_2$ molecules across all five hardware noise models. We measured the VQE accuracy ratio, gate count ratio, and circuit depth ratio before and after optimization, with the results summarized in Table \ref{tab:image_table}.

\begin{table}[t]
\centering
\caption{Impact of post-search optimization on VQE accuracy and circuit complexity ratios.}
\includegraphics[width=0.49\textwidth]{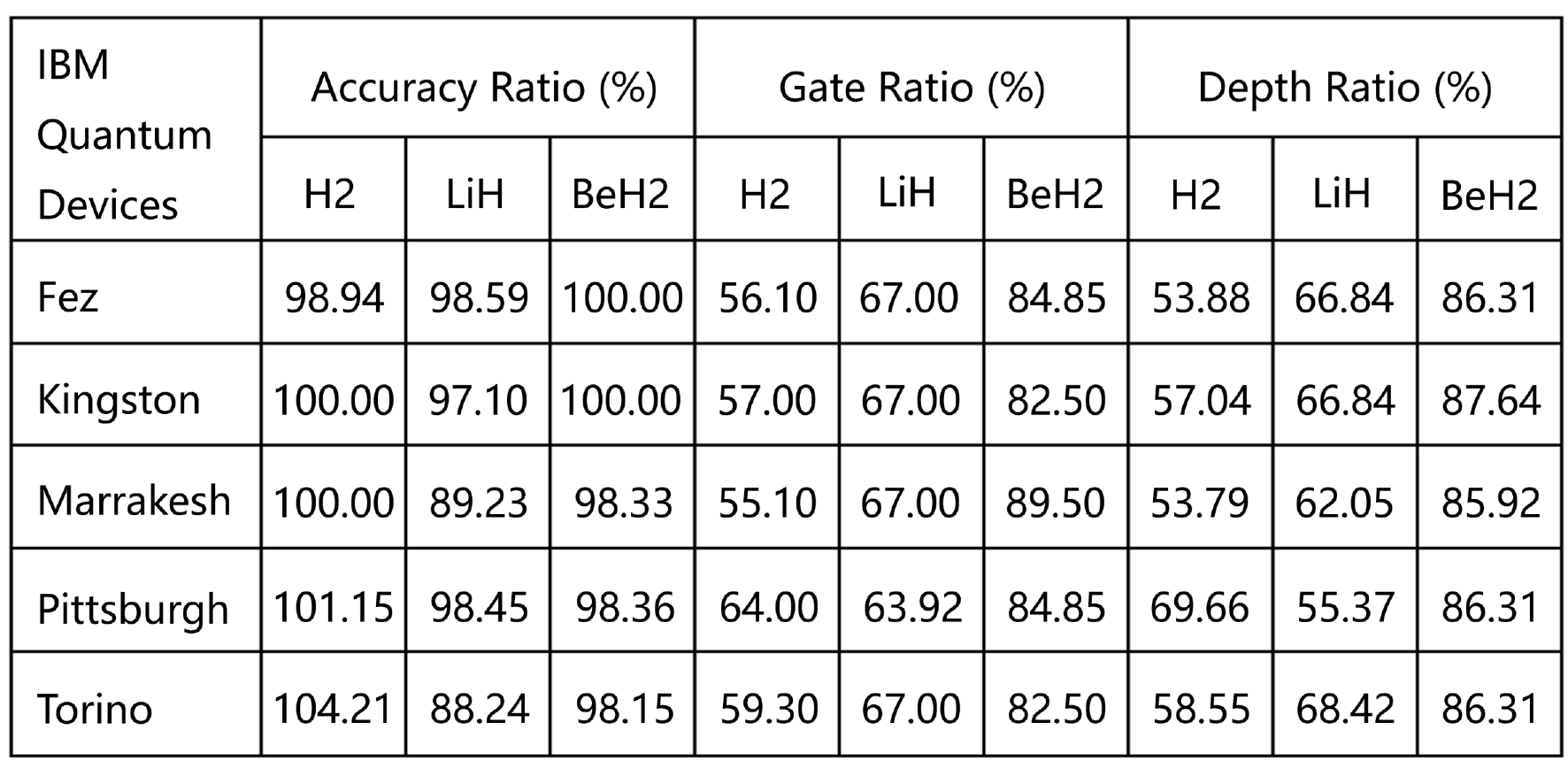}
\label{tab:image_table}
\end{table}

The data revealed that the optimization routine achieved substantial reductions in circuit complexity. For instance, for $H_2$ under the IBM Fez hardware noise model, gate count and depth were reduced to just 56.10\% and 53.88\% of their original values, respectively, with similarly significant reductions observed across all test cases. Critically, this level of simplification was achieved with only minimal degradation to the final VQE accuracy. The accuracy ratio was largely preserved, remaining close to 100\% in most scenarios and even exceeding it in some instances, with a peak of 104.21\% observed for $H_2$ under the IBM Torino hardware noise model. These results demonstrate a beneficial trade-off between circuit complexity and performance, as the substantial reduction in circuit size enhanced noise resilience. The accuracy improvement in noisy environments confirms that shallower, more compact circuits were less susceptible to hardware noise, validating our post-search optimization as an essential component for enhancing the viability of discovered circuits on near-term quantum processors.

\subsection{Large-Scale WSNs Routing Optimization}

\subsubsection{Problem Formulation and Modeling}
\begin{figure*}[t]
    \centering
    \includegraphics[width=0.9\textwidth]{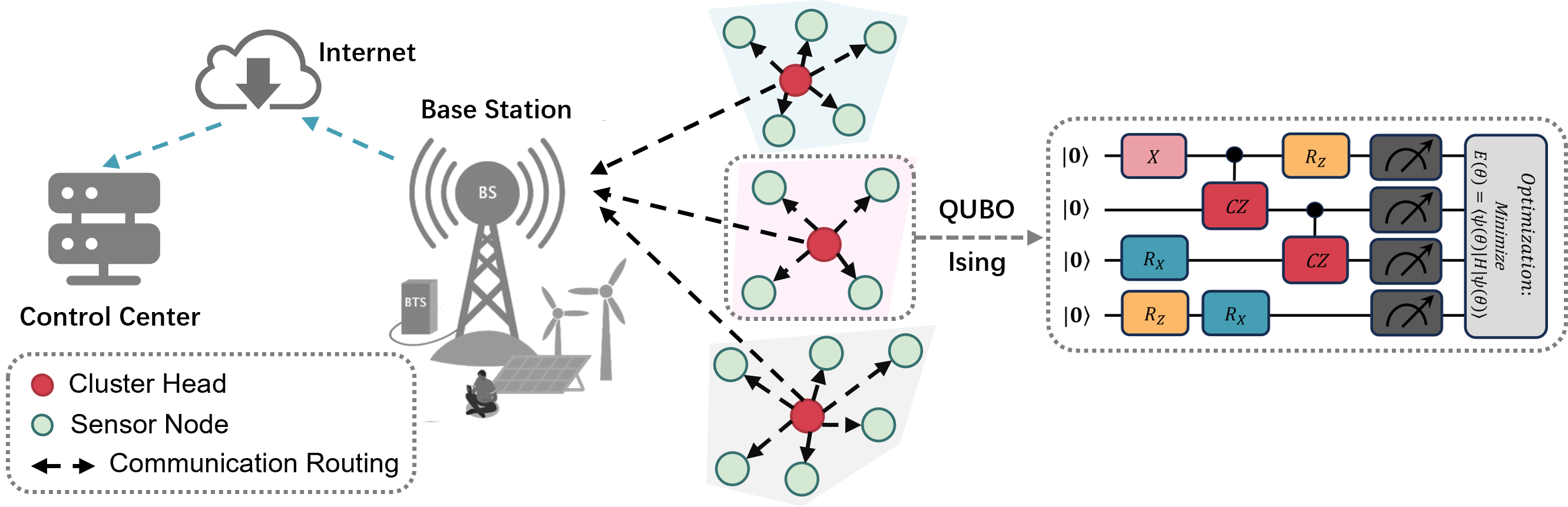} 
    \caption{Hybrid quantum-classical workflow for WSN routing optimization.} 
    \label{WSN}
\end{figure*}

This study investigates WSNs routing optimization, an NP-hard problem in distributed systems\cite{nguyen2018mobile}. A typical WSN comprises a large number of spatially distributed sensor nodes, which are responsible for collaboratively relaying collected data to a central base station. The primary operational objective is to maximize network longevity through an energy-aware routing protocol\cite{kocakulak2017overview}. This optimization is governed by two fundamental constraints: the finite battery capacity of each node and the restricted communication range. The restricted communication range necessitates data relaying through intermediate nodes, creating a complex combinatorial landscape, where suboptimal path selection can induce bottlenecks, leading to accelerated energy depletion, network partitioning, and catastrophic operational failure. While the NP-hard nature of this problem challenges classical algorithms at scale, its inherent network structure is well-suited for a hybrid quantum-classical decomposition, which mitigates scalability limitations by solving smaller, partitioned subproblems with quantum algorithms. An overview of this hybrid workflow is presented in Figure \ref{WSN}.

The WSN is modeled as a directed graph $G=(V,E)$, where $V$ is the set of nodes and $E$ represents the communication links. The node set $V$ is partitioned into three roles: sensors (Sensor), cluster heads (CH), and a single base station (BS). Each node $i \in V$ is assigned an initial energy $E_i$ corresponding to its function, with the energy for Sensors set to 100 units, CHs set to 200 units, and the BS provided with a functionally infinite supply. A directed edge $(i,j) \in E$ exists if the distance between nodes $i$ and $j$ is within a predefined maximum communication range $R$. The energy cost $c_{ij}$ of a transmission is modeled using the free-space path loss model, which is proportional to the squared Euclidean distance: 
\begin{equation} 
c_{ij} = \varepsilon \cdot \left( (x_i - x_j)^2 + (y_i - y_j)^2 \right)
\end{equation} 
where $(x_i, y_i)$ are the coordinates of node $i$ and $\varepsilon$ is an energy consumption coefficient.

To manage the problem's scale and align with the constraints of current quantum hardware, the network is partitioned into $k$ subgraphs, $G_s=(V_s,E_s)$, using spectral clustering\cite{2004A}. The routing task within each subgraph is then formulated as a Quadratic Unconstrained Binary Optimization (QUBO) problem\cite{lewis2017quadratic}. A binary variable $x_{ij} \in \{0,1\}$ is assigned to each edge $(i,j) \in E_s$, indicating its inclusion in the routing path.

The objective is to find a binary assignment that minimizes the total energy cost, subject to constraints on network flow conservation and node energy capacity. Flow conservation for each node $i$ is defined by the following constraint: \begin{equation} 
\sum_{j:(i,j)\in E_{s}}x_{ij}-\sum_{j:(j,i)\in E_{s}}x_{ji}=b_{i}
\end{equation} 
where $b_i$ represents the net data flow. The energy constraint ensures that the total transmission cost from any node $i$ does not exceed its initial energy capacity $E_i$: \begin{equation} 
\sum_{j:(i,j)\in E_{s}}c_{ij}x_{ij} \leq E_{i}
\end{equation} 
These constraints are incorporated as weighted penalty terms into the primary cost function, forming the final QUBO objective: 
\begin{equation}
\begin{split}
    \text{Minimize} \quad & \sum_{(i,j)\in E_{s}} c_{ij}x_{ij} + \lambda_{\text{flow}}\cdot\text{FlowConstraints}(\mathbf{x}) \\
    &+ \lambda_{\text{energy}}\cdot\text{EnergyConstraints}(\mathbf{x})
\end{split}
\end{equation}

The QUBO for each subgraph is solved using a VQA, which first maps the objective function to an Ising Hamiltonian $H_P$, whose ground state corresponds to the optimal routing solution. A PQC is then used to prepare the trial state $\ket{\psi(\boldsymbol{\theta})}$, and a classical optimizer tunes the parameters $\boldsymbol{\theta}$ to minimize the energy expectation value. This quantum-optimized intra-cluster routing is subsequently integrated with a classically computed inter-cluster backbone to form a globally connected solution. For this task, we employed our QBSA-DQAS framework to discover a problem-specific ansatz, and benchmarked the performance of the discovered ansatz against the standard QAOA ansatz\cite{chen2025resource}.

\subsubsection{Results and Analysis}
Following the problem formulation, we conducted the routing optimization on a large-scale WSN simulation. The network consisted of 109 nodes, including 100 Sensors, 8 CHs, and 1 BS, deployed within a $140 \times 70$ square units area. A communication radius of $R=25$ units resulted in an initial topology with 1750 links, which was partitioned into 5 clusters to ensure computational feasibility. Our QBSA-DQAS framework employed a $\{H, R_x, R_z, R_{zz}\}$ gate set for this task.

\begin{figure}[h]
    \centering
    \includegraphics[width=0.49\textwidth]{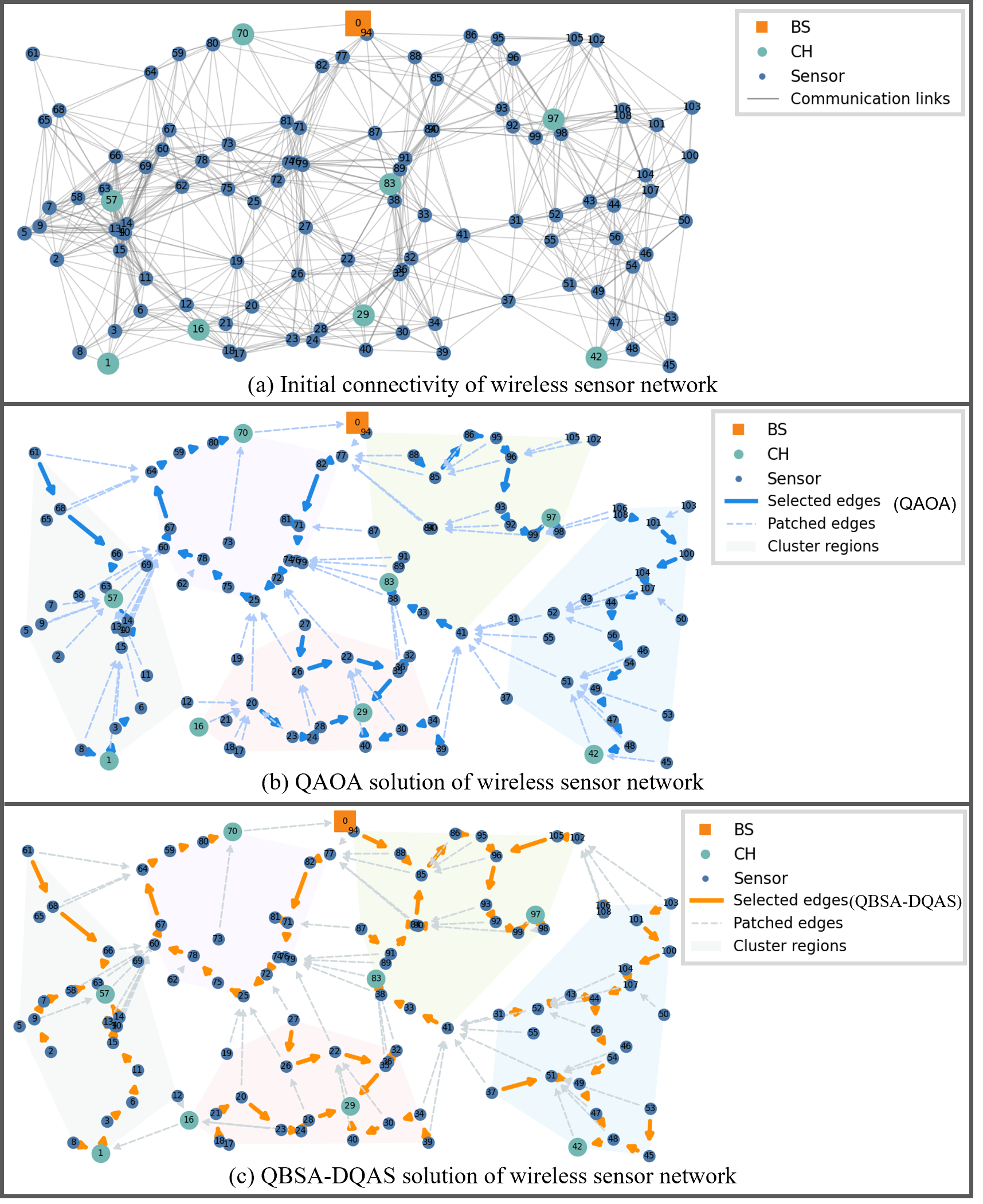} 
    \caption{Comparison of WSN routing solutions.} 
    \label{WSNres}
\end{figure}

The QBSA-DQAS approach yielded significant energy savings compared to both the standard QAOA and the classical greedy method. The results are presented in Figure \ref{WSNres}, which displays the initial network topology alongside the final optimized topologies. Quantitatively, our proposed method achieved a total energy cost of 2771.01 units, which was lower than the 3030.07 units of the QAOA solution and the 4671.78 units of the greedy algorithm, corresponding to a 40.7\% energy reduction over the classical greedy baseline and an 8.6\% improvement over the standard QAOA.

The optimized topology from QBSA-DQAS exhibits more coherent intra-cluster routing paths that form efficient, hierarchical routing structures. This superior structural integrity minimizes reliance on high-cost patched links required to ensure global connectivity. This result demonstrates the practical value of our architectural search for enhancing energy efficiency in large-scale WSN routing.

\section{Conclusion}
In this work, we have presented and experimentally validated the QBSA-DQAS framework, a novel approach designed to address the classical-quantum mismatch that hinders automated circuit design in the NISQ era. Our framework directly mitigates this issue by reframing architecture search as a meta-learning problem, where the QBSA module, employing learned quantum feature maps, natively captures the complex dependencies within the quantum system. We have further demonstrated that guiding this quantum-native search with a multi-objective function is a highly effective strategy. This function synergistically balances circuit expressibility and noise resilience to facilitate the discovery of powerful and practical architectures. The inclusion of a final post-search optimization stage, which minimizes circuit depth and gate counts, enhances the framework's practical utility by ensuring the generated circuits are tailored for near-term hardware.

Experimental evaluations across two computational domains have demonstrated the framework's practical utility and performance. For quantum simulation, circuits discovered for molecular ground-state energy estimation not only outperformed a suite of baselines in accuracy but also exhibited significant robustness across diverse hardware noise models. For combinatorial optimization, the ansatz generated for a large-scale WSN routing problem yielded a solution with substantially lower energy consumption, underscoring the framework's applicability to practical problems. These results collectively indicate that our integrated approach is both powerful and versatile. As a future direction, exploring the transferability and generalization capabilities of the QBSA-DQAS framework across different classes of computational problems may further enhance its practicality as a meta-learning tool. 
\begin{acknowledgments}
This work is supported by the National Natural Science Foundation of China (Grant No. 62471126), the Jiangsu Frontier Technology Research and Development Plan (Grant No. BF2025066), and the Fundamental Research Funds for the Central Universities (Grant No. 2242022k60001).
\end{acknowledgments}
\appendix
\section{Positional Encoding}
Sinusoidal positional encoding is defined as:
\begin{equation}
\begin{aligned}
\text{PE}(pos, 2i)   = \sin\left(pos/10000^{2i/d_{\text{model}}}\right)\\
\text{PE}(pos, 2i+1) = \cos\left(pos/10000^{2i/d_{\text{model}}}\right)
\end{aligned}
\end{equation}
where $pos$ is the depth position, $i$ is the dimension index, and $d_{\text{model}}$ is the model dimension. This enables the attention mechanism to capture sequential dependencies for optimal gate placement.

\section{Multi-Head Integration}
The outputs from all $N_h$ heads are concatenated and projected:
\begin{align}
\widetilde{Y} = \text{Concat}(Y^{(1)}, \ldots, Y^{(N_h)}) W_o + b_o
\end{align}
where $W_o$ and $b_o$ are output transformation parameters. This output is integrated with the input via residual connection:
\begin{align}
\alpha_{\mathrm{mid}} = \mathrm{LayerNorm}(\alpha_{\mathrm{in}} + \widetilde{Y})
\end{align}
producing the intermediate representation for subsequent position-wise transformation.

\bibliography{main.bib}

\end{document}